\documentclass[useAMS,usenatbib,twocolumn]{mn2e} 
\usepackage{graphicx,natbib}
\usepackage{url}
\usepackage{color}
\usepackage{times}
\usepackage{amssymb,amsmath}
\newcommand{\fsky}{f_{\rm sky}}
\newcommand{\nhat}{{\bf \hat n}}

\newcommand{\dbestbold}{{\bf d_{best}}}

\everymath{\displaystyle}

\begin{document}

\title{Dipolar modulation in number counts of WISE-2MASS sources}

\author[Mijin Yoon et al.]
{Mijin Yoon$^{1}$\thanks{{\texttt mijin@umich.edu}},
Dragan Huterer$^{1}$,
Cameron Gibelyou$^{1}$,
Andr\'as Kov\'acs$^{2}$,
and Istv\'an Szapudi$^{3}$
\\
${}^{1}$Department of Physics, University of Michigan, 
450 Church St, Ann Arbor, MI 48109-1040\\
${}^{2}$Institute of Physics, E\"{o}tv\"{o}s Lor\'and University, 1117 P\'azm\'any P\'eter s\'et\'any 1/A, Budapest, Hungary\\MTA-ELTE EIRSA "Lend\"ulet" Astrophysics Research Group, 1117 P\'azm\'any P\'eter s\'et\'any 1/A Budapest, Hungary\\
${}^{3}$Institute for Astronomy, University of Hawaii 2680 Woodlawn Drive, Honolulu, HI, 96822
}
\date{\today}
             
\maketitle

\begin{abstract}
We test the statistical isotropy of the universe by analyzing the distribution
of WISE extragalactic sources that were also observed by 2MASS. We pay
particular attention to color cuts and foreground marginalization in order to
cull a uniform sample of extragalactic objects and avoid stars. We detect a
dipole gradient in the number-counts with an amplitude of $\sim$0.05, somewhat
larger than expectations based on local structures corresponding to the depth
and (independently measured) bias of our WISE-2MASS sources. The direction of
the dipole, $(l,b)\simeq (310\,^{\circ}, -15\,^{\circ})$, is in reasonably good
agreement with that found previously in the (shallower) 2MASS Extended Source
Catalog alone. Interestingly, the dipole direction is not far from the
direction of the dipolar modulation in the CMB found by Planck, and also
fairly closely matches large-scale-structure bulk-flow directions found by
various groups using galaxies and type Ia supernovae.  It is difficult,
however, to draw specific conclusions from the near-agreement of these
directions.
\end{abstract}

\section{Introduction}\label{sec:intro}

Modern surveys of large-scale structure allow tests of some of the most
fundamental properties of the universe -- in particular, its statistical
isotropy. One of the most fundamental such tests is measuring the dipole in
the distribution of extragalactic sources. One expects a nonzero amplitude
consistent with the fluctuations in structure due to the finite depth of the
survey; this ``local-structure dipole'' in the nomenclature of
\citet{Gibelyou2012} is of order 0.1 for shallow surveys extending to $z_{\rm
  max}\sim 0.1$, but significantly smaller ($A\lesssim 0.01$) for deeper
surveys. The motion of our Galaxy through the cosmic microwave background
(CMB) rest frame also contributes to the dipole, but only at the level of
$v/c\simeq 0.001$; while this kinematic dipole was detected in the CMB a
long time ago, and more recently even solely via its effects on the higher
multipoles in the CMB fluctuations \citep{Aghanim:2013suk}, it has not yet been
seen in large-scale-structure (LSS) surveys.

Measurements of the dipole in LSS therefore represent consistency tests of the
fundamental cosmological model, and have in the past been applied to the
distribution of sources in NVSS
\citep{blake2002detection,Hirata2009,Rubart:2013tx,Fernandez-Cobos}.
Detection of an anomalously large (or small) dipole in LSS could indicate new
physics: for example, motion between the CMB and LSS rest frames, or the
presence of superhorizon fluctuations
\citep{zibin2008gauging,itoh2010dipole}. Moreover, in recent years,
measurements of the bulk motion of nearby structures have been conducted, out
to several hundred megaparsecs, using CMB-LSS correlations
\citep{kashlinsky2008measurement}, or out to somewhat smaller distances, using
peculiar velocities \citep{Watkins_2009,Feld_Watk_Hudson_2010}.

In this study, for the first time we test statistical isotropy using WISE
(Wide-field Infrared Survey Explorer) \citep{2010AJ....140.1868W}. WISE is, at
least at first glance, perfectly suited to tests of statistical isotropy since
it is deep and covers nearly the full sky. Moreover, its selection functions
have been increasingly well understood over the past few years based on its
observations in four bands sensitive to 3.4, 4.6, 12, and 22 $\mu m $
wavelengths with resolution in the 6"-12" range \citep{Yan:2012yk,
  Menard:2013aaa}.

\section{Culling of the WISE dataset} \label{sec:data}
Our measurement of the dipole relies on a suitable selection of a
representative sample of sources. The most important goal is to exclude
Galactic sources -- mainly stars. Galactic sources are expected to be
concentrated around the Galactic plane, with density falling off to the north
and south. While they are therefore expected to look like a $Y_{20}$ {\it
  quadrupole} in Galactic coordinates, the residual contamination of the
dipole may still be significant. Hence, in what follows we pay particular
attention to magnitude and color cuts applied to WISE in order to leave a
trustworthy set of extragalactic sources.

\begin{figure*}
\includegraphics[width=0.8\textwidth]{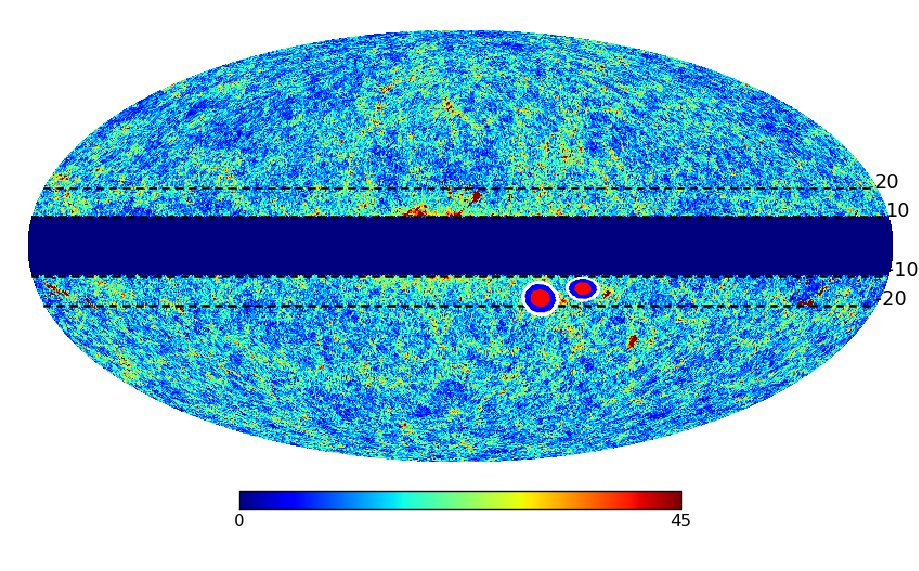}
\caption{Map of WISE-2MASS sources that we used with 10 degree Galactic cut (before masking out the contaminated
  region with the WMAP dust mask). The criteria are described in the text. The
  map shown is a Mollweide projection in Galactic coordinates with counts
  binned in pixels of about 0.5$ ^\circ$ on a side (HEALPix resolution ${\rm
    NSIDE}=128$). The two elliptical sets of contours represent the measured
  dipole direction when we applied a $10^{\circ}$ (left) and $20^{\circ}$
  (right) Galactic cut, respectively (that is, with $|b| < 10^{\circ}$ and
  $|b| < 20^{\circ}$). The red, blue, and white colors in those contours
  represent the $68\%$, $95\%$, and $99\%$ confidence regions for the
  direction.  }
\label{fig:map}
\end{figure*}

The Nov. 2013 release of WISE data includes 747 million objects in total.
Individual objects were not identified in the raw data, so data selection is
the key part of the analysis. We therefore apply carefully chosen criteria to
define a map as uncontaminated by Galactic objects as possible. As argued in
\citet{Kovacs:2013cga}, color cuts using only the WISE bands are not
sufficient, so we have applied 2MASS\footnote{Two Micron All Sky Survey
  \citep{skrutskie2006two}.} magnitudes ($J_{\rm 2mass}$) to distinguish
between stars and galaxies.  In other words, every source we use is observed
in both WISE and 2MASS, though we refer to our sample as ``WISE'' because
using that survey is crucial to give our sample greater depth.  To cull a
uniform, extragalactic sample of sources, we adopt the following color cuts:
\begin{itemize}

\item $W1 <15.2$, \\[-0.3cm]
\item $J_{\rm 2mass} < 16.5$,\\[-0.3cm]
\item $W1 - J_{\rm 2mass} < -1.7$. \\[-0.3cm]

\end{itemize}

Note that the first two criteria simply remove the faintest objects in the
respective band. To account for the effects of extinction by dust, 
  we correct the magnitudes for these two cuts using the SFD \citep{SFD}
  map\footnote{http://lambda.gsfc.nasa.gov/product/foreground/fg\_sfd\_get.cfm}. The third criterion above
represents the color cut that serves to separate galaxies from stars. The
detailed analysis on the data selection was described in
\cite{Kovacs:2013cga}; the resulting WISE map is shown in Fig.~\ref{fig:map}.

Unlike the previous studies that used WISE for cosmological tests
\citep{Kovacs:2013rs,Ferraro:2014msa}, our map does not show obvious
contamination in regions affected by the appearance of the Moon. Therefore, we
do not need to make further (and typically severe) cuts that remove these
regions. We do use the WMAP dust map \citep{Bennett:2012zja} to mask out the
pixels with remaining contamination; these mostly fall within $\pm 15^{\circ}$
Galactic latitude.  In addition, we cut out all pixels with $E(B-V)>0.5$
  from the SFD map (most of these have already been excluded by the
  WMAP dust map). We also checked for any unusual gradients with Galactic
latitude, especially around the Galactic plane, due to contamination from
stars.  These tests were consistent with zero gradient.

In the analysis, there are of order 2 million galaxies. We used the GAMA DR2
\citep{2008AAONw.114....3D} catalog to find sources in the WISE dataset that
are within 3'' of GAMA sources. We can thus determine the redshift
distribution of our objects. In the 144 sq.\ deg.\ overlapping region on the
sky, the matching rate is 96.9\%. The redshift distribution of matched
objects, $N(z)$, is shown in Figure \ref{fig:N(z)}; the mean is
$\bar{z}=0.139$. We use a smooth fit to the full distribution to obtain our
theoretical expectation for the local-structure dipole below.

\section{Methodology}

\subsection{Dipole estimator}
A robust and easy-to-implement dipole estimator was first suggested by
\citet{Hirata2009}, who measured hemispherical anomalies of quasars, and later
adopted by \citet{Gibelyou2012} to measure the dipole in a variety of LSS
surveys. The number of sources in direction $\hat{\mathbf{n}}$ can be written as
\begin{equation}
N(\hat{\mathbf{n}})=[1+A\,\hat{\mathbf{d}}\cdot\hat{\mathbf{n}} ]{\bar N }+\epsilon (\hat{\mathbf{n}})
\end{equation}
where $A$ and $ \hat{\mathbf{d}}$ are the amplitude and direction of the
dipole, and $\epsilon$ is noise.  

The modulation in number counts can be written as the sum of contributions
from a dipole, fluctuations due to systematics, and a mean offset
\citep{Hirata2009}.
\begin{equation}
\delta N/\bar{N} = A\, {\bf \hat d} \cdot {\bf \hat n} + \sum_i k_i t_i({\bf \hat n}) + C.
\label{eq:N_with_templates}
\end{equation}
Here $t_i({\bf \hat n})$ represent the systematics maps, while the
coefficients $k_i$ give the amplitudes of the contributions of these
systematics to the observed density field. The presence of the monopole term,
$C$, allows us to account for covariance between the monopole and other
estimated parameters, especially covariance between the monopole and any
systematic templates. The best linear unbiased estimator of the combination
({\bf d}, $k_i$, $C$), with corresponding errors, is obtained as
follows. First, we rewrite the above equation as $\delta N/N = {\bf x} \cdot {\bf
  T(\hat n)}$ where ${\bf x} = (d_x, d_y, d_z, k_1, ..., k_N, C)$, ${\bf
  T(\hat n)} = (n_x, n_y, n_z, t_1(\hat n), ..., t_N(\hat n), 1)$, and $n_x^2
+ n_y^2 + n_z^2 = 1$.  The best linear unbiased estimator of {\bf x} is
\begin{equation}
{\bf \hat x} = F^{-1} g
\end{equation}
where the components of the vector $g$ are $g_i = \int T_i(\hat n) \delta
N^{\Omega}(\hat n) d^2 \hat n$ and the Fisher matrix $F$ is given by $F_{ij} =
\bar N^{\Omega} \int T_i(\hat n) T_j(\hat n) d^2 \hat n$, where $N^{\Omega}
\equiv dN/d\Omega$ is the number of galaxies per steradian ($\Omega$ is a
solid angle). The integrals from
which the vector $g$ and the Fisher matrix $F$ are calculated are discretized
in our survey. We adopt a HEALPix \cite{healpix} pixelization with {\tt
  NSIDE}=128, so that each pixel corresponds to about half a degree on a
side and contains roughly 14 sources.

The formalism above returns the best-fit dipole components (first three
elements of the vector ${\bf x}$), together with their covariance (inverse of
the corresponding Fisher matrix). We are however most interested in the
likelihood of the amplitude of the dipole, $A=(d_x^2+d_y^2+d_z^2)^{1/2}$. We
can construct a marginalized likelihood function for the amplitude $A$
\citep{Hirata2009}:
\begin{equation}
\mathcal{L}(A) \propto \int \exp \left \lbrack -\frac{1}{2}(A \nhat -
\dbestbold) {\rm Cov}^{-1}(A \nhat - \dbestbold) \right \rbrack d^2 \nhat
\label{eq:like_A}
\end{equation}
where $d^2 \nhat$ indicates integration over all possible directions on the
sphere. Thus we readily obtain a full likelihood for the amplitude. In our
results, we quote the $68\%$ region around the best-fit amplitude.


\begin{figure}
\includegraphics[width=0.4\textwidth]{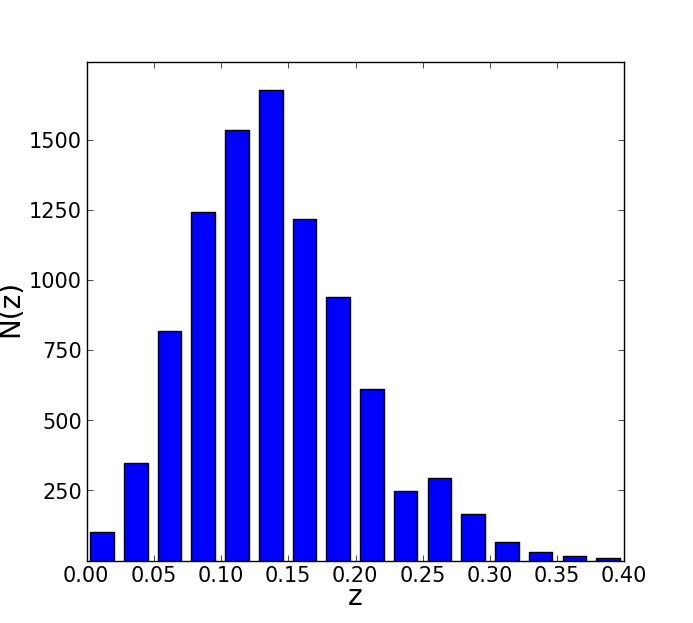}
\caption{Number counts of WISE sources as a function of redshift. We obtain
  redshift information by matching WISE sources to those from the GAMA DR2
  catalog. As explained in the text, matching works very well.}
\label{fig:N(z)}
\end{figure}

\subsection{Foreground Templates and Estimator Validation}

Despite our carefully chosen magnitude and color cuts, it is likely that there
is some star contamination to our extragalactic source map. Moreover, on a cut
sky, the dipole is not completely decoupled from the monopole, quadrupole, and
other multipoles, and hence we need to marginalize over some of them in order
to get correct results. We therefore include several templates -- maps
$t_i({\bf \hat n})$ in the parlance of Eq.~(\ref{eq:N_with_templates}) -- with
amplitudes $k_i$ over which we marginalize:
\begin{itemize}
\item To deal with the remaining star contamination, we add a star map as a
  template.  The star map was generated based on the Tycho 2 catalog
  \citep{2000A&A...355L..27H}, as suggested in \cite{Kovacs:2013rs}. The
  inclusion of this template affects the measured dipole negligibly,
  reinforcing our confidence that star contamination does not affect the
  result.
\item To account for the other multipoles, we add the monopole (corresponding
  to the constant $C$ in Eq.~(\ref{eq:N_with_templates}) with no spatial
  dependence), as well as the quadrupole and octopole that include 5 and 7
  extra parameters. We therefore marginalize over these 13 parameters in
  addition to the amplitude of the star map.  We experimented with
  marginalization over a few more ($\ell\geq 4$) multipoles, but for small
  Galactic cuts ($b_{\rm cut}\lesssim 15^\circ$), the shift in the dipole
  direction and magnitude were small.
\end{itemize}

We validated our estimator by running simulations with an input dipole of a
given amplitude assuming various sky cuts and marginalizing over templates. We
verified that the input dipole is recovered within the error bars.

\subsection{Theoretical expectation}

We calculate the theoretical expectation for the local-structure dipole using
standard methods (see e.g.\ Sec.\ 2.2 of \citet{Gibelyou2012}). We calculate
the angular power spectrum of large-scale structure for the given source
distribution $N(z)$, and evaluate it at the dipole ($C_\ell$ at $\ell=1$);
this calculation does not assume the Limber approximation since the latter is
inaccurate at these very large scales. The amplitude is then given as $A_{\rm
  theory}=(9C_1/(4\pi))^{1/2}$ \citep{Gibelyou2012}, while the theory error is
given by cosmic variance for $\ell=1$: $\delta A_{\rm theory}/A_{\rm
  theory}=(1/2)\sqrt{2/((2\ell+1)\fsky)}=(6\fsky)^{-1/2}$.  Evaluating the
theoretically expected dipole for the source distribution shown in
Fig.~\ref{fig:N(z)}, we get
\begin{equation}
A_{\rm theory} = (0.0233 \pm 0.0094\fsky^{-1/2})\times 
\left (\frac{{\rm bias}}{1.41}\right )
\label{eq:A_th}
\end {equation}
Here we make explicit the dependence of the cosmic variance error on the
fraction of the sky covered $\fsky$, and also on the bias of WISE sources. To
obtain the latter, we followed \citet{Kovacs:2013cga}, and estimated the bias
of the galaxy catalog using {\tt SpICE} \citep{spice} and the Python 
CosmoPy\footnote{\texttt{http://www.ifa.hawaii.edu/cosmopy/}} package. We note
that the estimation of the bias is particularly sensitive to $\sigma_{8}$
because they both act to renormalize the angular power spectrum, and in linear
theory $C^{gg}_{\ell} \propto (b\sigma_{8})^{2}$. We fix $\sigma_{8}=0.8$ in
our measurements, finding $b=1.41\pm0.07$. This value is comparable to earlier
findings \citep{RassatEtal2003} that measured a value of $b=1.40\pm0.03$ for
a 2MASS selected galaxy sample.

\begin{figure}
\includegraphics[width=0.48\textwidth]{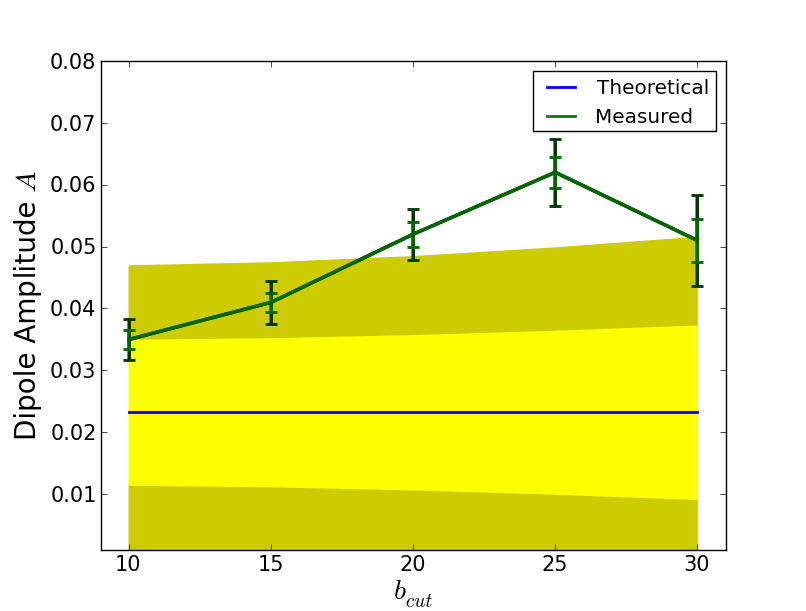}
\caption{Theoretical prediction for the dipole amplitude
  (horizontal blue line), together with the measured values in WISE (green
  points). The two sets of error bars on the measurements correspond to 68\%
  and 95\% confidence; they have been calculated from the full likelihood in
  Eq.~(\ref{eq:like_A}) and are rather symmetric around the maximum-likelihood
  value. The two large horizontal bands around the theory prediction
  correspond to 1- and 2-sigma cosmic variance error.  }
\label{fig:A_th_and_directions}
\end{figure}

\section{Results}

Our measurements of the dipole's amplitude and direction, as a function of the
(isolatitude) Galactic cut, are presented in Table \ref{tab:A}. The best-fit
direction of the dipole is also shown in Fig.~\ref{fig:map} for the $10^\circ$
and $20^\circ$ Galactic cut, the two cases roughly illustrating the dependence
of the direction on the Galactic cut.

We first note a reasonably good consistency between the recovered
directions, despite the fact that the number of sources decreases by a factor
of $\sim$1.4 as we increase the Galactic cut in the range shown. We also note
that the overall amplitude is roughly 1.5 - 2.7 times larger than the
theoretically expected one, and is roughly 1-2$\sigma$ high, where $\sigma$
corresponds to cosmic variance since the measurement error is much smaller
(see Table \ref{tab:A}). Finally, we note that while the dipole amplitude does
vary with $b_{\rm cut}$ more than its typical measurement errors, it is
overall consistent at $A_{\rm WISE}\simeq 0.04$-$0.05$, which is rather
robustly stable given the large decrease of the number of sources with
increasing Galactic cut.

It is interesting to note that 2MASS Extended Source Catalog data, as analyzed in \citet{Gibelyou2012}
(redshift $0<z<0.2$, $N =3.8\times10^5$), give $A_{\rm 2MASS} = 0.104 \pm
0.004$, $(l,b) = (268.4^{\circ}, 0.0^{\circ})$ -- amplitude higher than ours
due to the greater contribution of the local-structure dipole for the
shallower survey, direction not far. Relative to this previous work, we have
therefore made progress by pushing down a factor of 2.5 in the dipole
amplitude. This is a welcome development toward being able to probe the
kinematic dipole due to our motion relative to the overall LSS rest frame, which will
require reaching the level $A \sim 10^{-3}$, and therefore a deeper survey (or a
deeper sample of WISE sources).

\begin{table}\footnotesize
\setlength{\tabcolsep}{0.35em}
\begin{tabular}{|| c | c | c | c | c ||}
\hline 
\rule[-3mm]{0mm}{8mm} $b_{\rm cut}$& $\fsky$ & $A_{\rm WISE}$ & $A_{\rm theory}$& $\hat{\mathbf{d}}(l\,^{\circ},b\,^{\circ})$ \tabularnewline
\hline
\hline 
\rule[-3mm]{0mm}{8mm} $10^{\circ}$& 0.65 & $0.035\pm 0.002$ & $0.023 \pm
0.012$ & $(326\pm3, \,-17\pm2)$ \\\hline 
\rule[-3mm]{0mm}{8mm} $15^{\circ}$ & 0.62 & $0.042\pm 0.002$ & $0.023 \pm
0.012$ & $(316\pm3, \,-15\pm2)$ \\\hline 
\rule[-3mm]{0mm}{8mm} $20^{\circ}$& 0.57 & $0.052\pm 0.002$ & $0.023 \pm
0.012$ & $(308\pm4, \,-14 \pm2)$ \\\hline
\rule[-3mm]{0mm}{8mm} $25^{\circ}$& 0.51 & $0.062\pm 0.003$ & $0.023 \pm
0.013$ & $(315\pm6, \,-12 \pm2)$ \\\hline
\rule[-3mm]{0mm}{8mm} $30^{\circ}$& 0.45 & $0.051\pm 0.004$ & $0.023 \pm
0.014$ & $(335\pm6, \,-18 \pm3)$ \\\hline\hline
\end{tabular}
\caption{Measurements of the dipole amplitude in WISE for various Galactic
  cuts ($b_{\rm cut}$) corresponding to fractions of the sky covered
  ($\fsky$). In all cases we marginalized over several foreground templates,
  as described in the text. The full likelihood for the amplitude $A_{\rm
    WISE}$ is well approximated by a Gaussian whose mode and standard
  deviation we quote here. We also show the theoretical expectation $A_{\rm
    theory}$ due to the local-structure dipole, together with the
  corresponding cosmic variance given a bias $b=1.41$.}
\label{tab:A}
\end{table}

\section{Conclusions}\label{sec:concl}

We measured the clustering dipole in the WISE survey, using a carefully culled
sample that contains 2 million extragalactic sources with a known redshift
distribution.  The amplitude of the measured dipole is $A\simeq 0.05\pm
  0.01$, where we quote the central value corresponding to the $20^{\circ}$
  cut case and error that shows the dispersion of central values for
  $15^{\circ}\leq b_{\rm cut}\leq 25^{\circ}$. The amplitude is therefore
  roughly twice as large as the theoretical expectation; see
  Eq.~(\ref{eq:A_th}). The direction of the dipole is  $\simeq
  (310^\circ\pm5, \,-15^\circ\pm2)$. 

What could explain the excess dipole measured relative to theoretical
expectation? The systematics, while an obvious first suspect, are not
necessarily at fault given the rather extensive care we took to account for
them: we carefully culled the dataset by imposing cuts based on WISE and 2MASS magnitudes; we
included cuts based on Galactic latitude and on the WMAP dust map, and we
further marginalized over a carefully derived star-map template as well as
templates corresponding to the quadrupole and octopole.

Another possibility is that the excess signal is cosmological. For example, a
large void might generate the excess observed here
\citep{Rubart:2014lia}. Such a void was incidentally just detected in the
analysis of the WISE data itself \citep{Szapudi:2014zha,FinelliEtal2014}. At this time it
is too early to tell whether the WISE void is contributing significantly
to the excess dipole that we measured, though a rough comparison with numbers
in \citet{Rubart:2014lia} appears to indicate that it is not.

It is also interesting to note that Planck found a best-fit modulation with
both amplitude and direction roughly (within $\sim$3$\sigma$ of their errors)
in agreement with ours \citep{Ade:2013nlj}: $A_{\rm Planck} = 0.078 \pm 0.021$,
$(l,b) = (227^\circ, -15^\circ) \pm 19^\circ $.  It is not clear at this time
what, if any, significance to assign to the comparable-looking modulations in
WISE and Planck since their sources are at vastly different redshifts ($z\sim
0.15$ and $\sim 1000$), and the agreement in amplitude and direction is only
approximate. Finally, the direction we find is {\it also} close to the
peculiar-velocity bulk-flow directions found using type Ia supernovae
\citep{Dai_Kinney_Stojkovic,Kalus_SN,Rathaus_SN}, galaxies
\citep{Feld_Watk_Hudson_2010, Turnbull,Ma_Gordon_Feldman,Ma_Pan}, and the
kinetic Sunyaev-Zeldovich effect \citep{Lavaux_kSZ}. While the agreement
between the directions is suggestive, it is not immediately clear how our WISE
dipole is related to these. For example, interpreting the excess dipole
amplitude $\delta A\sim 0.03$ as a bulk motion is clearly out of the question,
since it would correspond to a huge velocity of $v\simeq 0.015c = 4500$ km/s,
an order of magnitude larger than what typical bulk-motion measurements
indicate.


With recent measurements of the cross-correlation of its sources with the
CMB and the detection of a large underdense void, WISE is finally making major
contributions to cosmology. Its nearly all-sky coverage is a huge asset and
gives the survey a big advantage on that front over most other LSS surveys. In this paper
we have taken another step in testing fundamental cosmology with WISE by
measuring the clustering dipole in the distribution of its extragalactic
sources. We look forward to further investigations of this result, especially
in conjunction with other related findings in the CMB and LSS.

\section*{Acknowledgements}
We thank Maciej Bilicki, Brice M\'{e}nard and Dominik Schwarz for
comments on the manuscript, and Matt Ashby and Jeffrey Newman for discussion
of color cuts. The work of MY and DH has been supported by the DOE and the
NSF. AK acknowledges OTKA grant no.\ 101666, and support from the Campus
Hungary fellowship program. IS acknowledges NASA grants NNX12AF83G and
NNX10AD53G. This work made use of the HEALPix package~\citep{healpix}.

\bibliographystyle{mn2e}
\bibliography{mnras_wise_dipole}

\end{document}